\documentclass[3p]{elsarticle}

\usepackage{amssymb}

\usepackage{amsmath}
\usepackage{caption}
\usepackage{subcaption}
\usepackage{multirow}

\usepackage{adjustbox}
\usepackage{hyperref}
\usepackage{stackengine}

\begin{document}

\begin{frontmatter}

\title{Scaling properties of extreme price fluctuations in Bitcoin markets}

\author{Stjepan Begu\v{s}i\'{c}\textsuperscript{1}\corref{cor1}, Zvonko Kostanj\v{c}ar\textsuperscript{1}, H. Eugene Stanley\textsuperscript{2}, and Boris Podobnik\textsuperscript{3,4,5} }

\address{\textsuperscript{1}Laboratory for Financial and Risk Analytics, Faculty of Electrical Engineering and Computing,\\ University of Zagreb, 10000 Zagreb, Croatia\\  \textsuperscript{2}{Center for Polymer Studies and Department of Physics, Boston University, Boston, Massachusetts 02215, USA}\\  \textsuperscript{3}{Faculty of Civil Engineering, University of Rijeka, 51000 Rijeka, Croatia} \\ \textsuperscript{4}{Zagreb School of Economics and Management, 10000 Zagreb, Croatia} \\ \textsuperscript{5}{Luxembourg School of Business, Luxembourg, Grand-Duchy of Luxembourg}}

\cortext[]{Corresponding author: Stjepan Begu\v{s}i\'{c} (\href{mailto:stjepan.begusic@fer.hr}{stjepan.begusic@fer.hr}).}

\begin{abstract}
Detection of power-law behavior and studies of scaling exponents uncover the characteristics of complexity in many real world phenomena. The complexity of financial markets has always presented challenging issues and provided interesting findings, such as the inverse cubic law in the tails of stock price fluctuation distributions. Motivated by the rise of novel digital assets based on blockchain technology, we study the distributions of cryptocurrency price fluctuations. We consider Bitcoin returns over various time intervals and from multiple digital exchanges, in order to investigate the existence of universal scaling behavior in the tails, and ascertain whether the scaling exponent supports the presence of a finite second moment. We provide empirical evidence on slowly decaying tails in the distributions of returns over multiple time intervals and different exchanges, corresponding to a power-law. We estimate the scaling exponent and find an asymptotic power-law behavior with $2< \alpha < 2.5$ suggesting that Bitcoin returns, in addition to being more volatile, also exhibit heavier tails than stocks, which are known to be around 3. Our results also imply the existence of a finite second moment, thus providing a fundamental basis for the usage of standard financial theories and covariance-based techniques in risk management and portfolio optimization scenarios.
\end{abstract}


\end{frontmatter}

\section{Introduction}

The dynamics of many ecological, social, technical or economic systems are known to exhibit distributions with a power-law decay in the tails \cite{Mantegna1995,Clauset2009}, which is generally a symptom of complexity and self organized criticality in dynamical systems. A prominent example are financial markets, for which the \emph{inverse cubic law} of stock price fluctuations is a well established fact \cite{Cont2001,Gabaix2003,Gabaix2009}. This finding states that the empirical cumulative distributions for extreme price changes over various time intervals, ranging from high--frequency to daily, decay with an exponent of $\alpha \simeq 3$, as a universal scaling law \cite{Gopikrishnan1998,Gopikrishnan1999,Plerou1999}. It also implies the existence of a finite second moment for price changes, which has direct consequences for risk management and portfolio optimization, where covariance--based methods are primarily used. The emergence of power-law behavior in price fluctuations is argued to be a consequence of underlying complex mechanisms, such as feedback effects and correlations in financial markets \cite{Gopikrishnan1998,Gopikrishnan1999,Podobnik2001,Podobnik2009}. Some theories associate this phenomenon with market impact and the distribution of large investors \cite{Gabaix2003,Gabaix2009}, while other studies model the power-law behavior as a consequence of investors' poor estimation of the true value of a company and limited information \cite{Kostanjcar2013}. While there are various models which can generate adequate statistical properties of stock prices \cite{Maslov2000}, the reported universal scaling behavior in price fluctuations is yet to be comprehensively studied across various financial markets \cite{Bouchaud2001,Farmer2004,Sornette2014}. 

The increased adoption of blockchain technology and the emergence of cryptocurrencies as a novel asset class led to the development of online cryptocurrency exchanges, mostly open 24/7 to investors worldwide. In this paper we focus on Bitcoin \cite{Nakamoto2008}, the largest cryptocurrency in terms of market capitalization, and the one with the longest history as a digital asset. In recent years we have witnessed both hysteria and panic materialize into several Bitcoin bubbles and crashes of substantial proportions from 2010 to today - something not usually seen in stock, currency or commodity markets. Understanding the distributional properties of cryptocurrency price fluctuations is a necessary step towards appropriate investment approaches to cryptocurrency markets \cite{Briere2015,Dyhrberg2016}. Osterrieder and Lorenz \cite{Osterrieder2017} recently presented a statistical analysis of the the extreme tail behavior using daily cryptocurrency price data, finding that Bitcoin returns exhibit volatility several times larger than that of large \emph{fiat} currencies, but also heavier tails as suggested by the fitted generalized Pareto and extreme value distributions. Other research efforts have been focused on modelling and explaining the associated phenomena stemming from the underlying complex system of miners and users in the blockchain and investor networks \cite{Kondor2014,ElBahrawy2017}. Information patterns in online digital and social networks were linked to price performance of Bitcoin and were found to exhibit feedback relationships with the Bitcoin market performance \cite{Kristoufek2013,Garcia2014}. This line of research was advanced by the identification of econometric evidence for the emergence of bubbles and crashes in Bitcoin, implying the speculative nature of Bitcoin early adopters \cite{Cheung2015}. Donier and Bouchaud \cite{Donier2015} found that the market microstructure on Bitcoin exchanges can be used to anticipate illiquidity issues in the market, which lead to abrupt crashes. Further research focused on identifying main drivers of Bitcoin prices in the market, finding that prices were driven mostly by endogenous fundamental factors \cite{Kristoufek2015,Ciaian2016}. Although many statistical features of cryptocurrency returns were analyzed \cite{Balcilar2017,Bariviera2017}, no evidence has been reported on the scaling properties in the disitrbution tails, or the existence of a universal behavior. 

In this paper we present a detailed study of the heavy tail properties in empirical Bitcoin return distributions. We inspect the cumulative distributions at extreme values and find a power-law decay for multiple time intervals and different exchanges. Since the power-law exponents are notoriously difficult to estimate, we employ two estimation methods and a resampling-based technique to statistically validate the power-law hypothesis in the observed distribution tails. We also check whether the evolution of liquidity in Bitcoin exchanges through time has had an effect on the scaling exponents. We find that the exponent $\alpha$ seems to lie within the range $2<\alpha<2.5$, which is at odds with the \emph{inverse cubic law} ($\alpha \simeq 3$) found for stocks \cite{Gopikrishnan1998,Gopikrishnan1999,Plerou1999}, meaning that in addition to much larger market volatility \cite{Dyhrberg2016}, Bitcoin returns also exhibit heavier tails than stocks \cite{Cont2001}. However, the fact that the exponent is found to be outside the Levy-stable region $0<\alpha < 2$ implies the existence of a finite second moment (i.e. variance), which is a key foundation for risk estimation methods and techniques, and ultimately for portfolio optimization scenarios which include Bitcoin as an investment.

\section{Data}
We analyze trade--level data of the BTC/USD pairs from five large Bitcoin exchanges: Mt. Gox, BTC-e, Bitstamp, Bitfinex, and Kraken. These digital exchange platforms were active throughout various time periods - the availability of data is shown in Table \ref{data_table}. Each data point consists of a trade price (Bitcoin price in U.S. dollars) and the traded quantity. 

\begin{table}[h]
\centering
\small
\bgroup
\def\arraystretch{1.2}
\begin{tabular}{ |r|c|c| }
\hline
     Exchange & Period & Number of trades \\ \hline \hline
     
     Bitfinex & Mar 2013 -- Dec 2016 & 10,182,187 \\ \hline
     Bitstamp & Sep 2011 -- Feb 2018 & 21,353,815 \\ \hline
     BTC-e & Aug 2011 -- Jul 2017 & 32,904,778 \\ \hline
     Kraken & Jan 2014 -- Feb 2018 & 7,635,562 \\ \hline
     Mt. Gox & Jul 2010 -- Feb 2014 & 8,295,806 \\ \hline

\end{tabular}
\egroup
    \caption{Summary for the considered Bitcoin exchange data.}
    \label{data_table}
\end{table}

\begin{figure}[h!]
\center
\includegraphics[width=0.75\columnwidth]{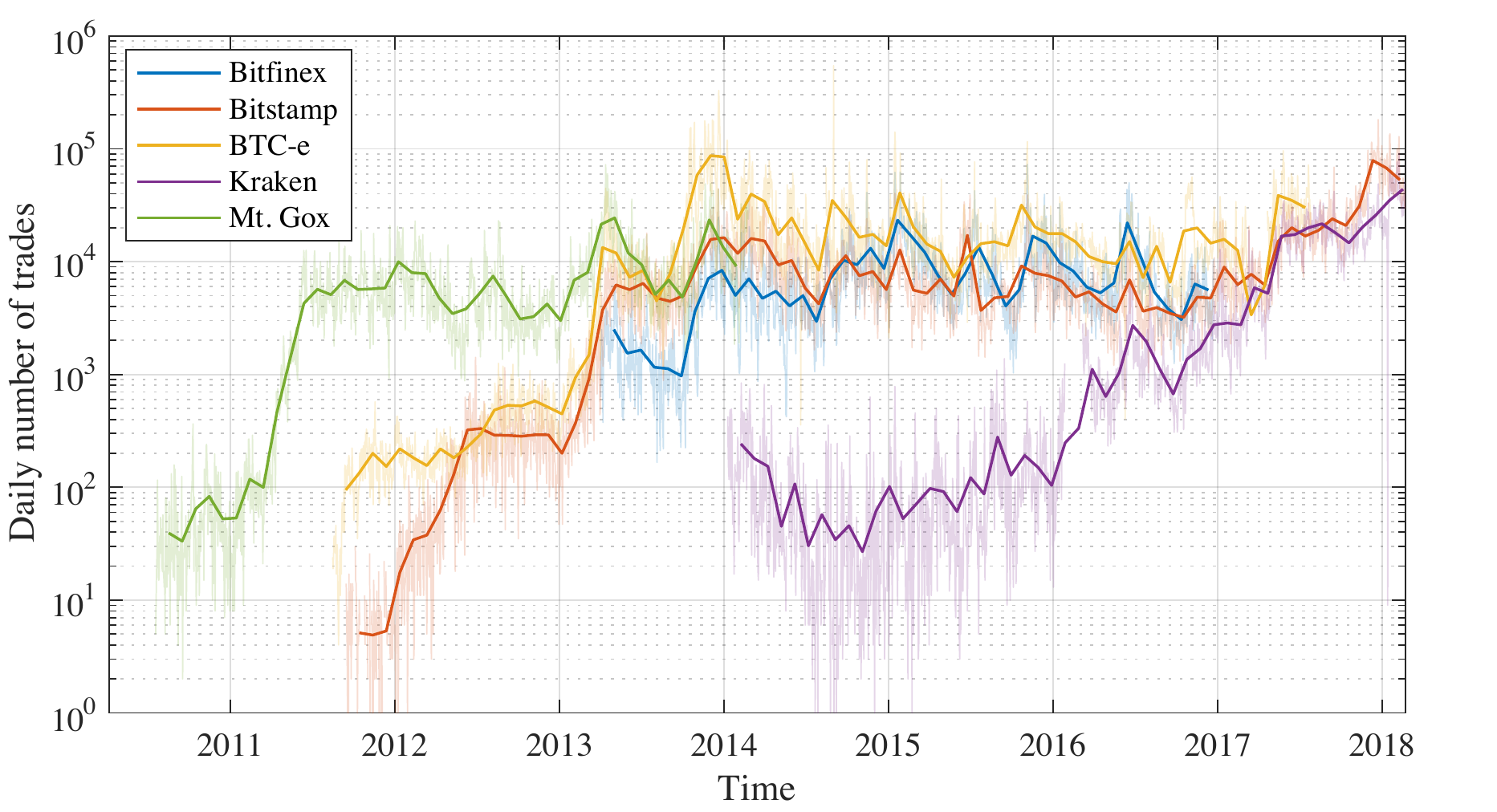}
\caption{Daily number of trades (transparent lines) and their 30-day moving average (solid lines) for the considered Bitcoin exchanges throughout the entire period of analysis, shown in logarithmic scale. The discontinued periods correspond to missing data in the datasets.}
\label{ntrades}
\end{figure}

\noindent Although the considered time period of 7 years is not particularly long when compared to analyses performed on stock data spanning several decades \cite{Gopikrishnan1998,Gopikrishnan1999,Plerou1999} (this is especially important for estimation of distribution tails), the fact that data from multiple exchanges is used introduces an additional dimension to our analysis. Firstly, different exchanges have been operational at different periods in time, and the persistence of the scaling properties of Bitcoin price fluctuation distributions can be studied by encompassing a period as large as possible. Moreover, the exchanges have different user bases, meaning that although we only consider the BTC/USD pairs, various effects of illiquidity may mean that the price fluctuation distributions for individual exchanges are not the same. As seen in Figure \ref{ntrades}, Bitcoin trading activity has undergone large shifts over the considered time period, from the 2013 boom and the subsequent stagnation, to the recent expansion in both trading activity and the number of exchanges throughout 2017. Apart from the apparent evolution in number of trades, we also report a similar behavior for the trading volume. It is worthy of noting that Bitcoin trading ecosystem also suffered some distress, such as the Mt. Gox theft and its liquidation in February 2014, and the BTC-e seizure in 2017 by the U.S. Department of Justice on allegations of money laundering and connection to the Mt. Gox hack. 

Using the trade--level data (a total of over 80 million individual trades) we calculate price fluctuations on various fixed time intervals, ranging from $\Delta t = 1$ min to $\Delta t = 1$ day. The price $P_t$ at time $t$ is calculated as the last traded price in the previous period from $t-\Delta t$ to $t$. Using the evenly--spaced price time series, we calculate the log--returns as the change in the logarithm of the price:

\begin{equation}
R_{t} = \log{P_{t}} - \log{P_{t-\Delta t}}.
\end{equation}

\noindent In our analysis the returns for each considered exchange and time interval are standardized (in order to measure how many standard deviations their values are from the mean) by subtracting the mean and dividing by the standard deviation:

\begin{equation}
R_{t}^{norm.} = \frac{R_{t} - \mu_{R}}{\sigma_{R}},
\end{equation}

\noindent where $\mu_{R}$ and $\sigma_{R}$ denote the mean and standard deviation of returns for the considered exchange. An additional advantage of using standardized returns is the ability to compare the estimated empirical distributions to each other across different time intervals and exchanges.

\section{Estimation of tail exponents in distributions of price fluctuations}
A well-known property of financial asset returns is the \emph{inverse cubic law}, which implies that the tails of their cumulative distributions follow a power-law:

\begin{equation}
P(X>x) \sim x^{-\alpha},
\label{PL}
\end{equation}

\noindent where the scaling exponent is $\alpha \simeq 3$ (the respective probability density functions tails thus decay with $\alpha + 1 \simeq 4$). This characteristic exponent is found in data from many different financial markets, including individual stocks in developed markets \cite{Plerou1999}, market indices \cite{Gopikrishnan1999}, as well as developing markets \cite{Pan2008}. The fact that the inverse cubic law is found on time scales ranging from $\Delta t = 1$ min to $\Delta t = 1$ day testifies to the universality of this scaling behavior.

Identifying and quantifying the tail properties of such distributions is generally not a trivial task. Specifically, to estimate the power-law exponent $\alpha$ first one needs to determine the lower bound $x_{min}$, which essentially isolates data believed to belong to the tail of the distribution. Then, the exponent can be inferred from the tail data using several techniques, among which the most popular are a regression fit in the log-log scale, and the Hill estimator \cite{Hill1975}. Plotting the tail distribution on a log-log scale has long been a popular method for the visual inspection of power-law behavior - thus the linear regression stems as a natural choice for the estimation of the scaling exponent. On the other hand, the Hill estimator is the maximum likelihood estimator under the assumption that the tail data follows a  power-law \cite{Clauset2009}, and reads: 

\begin{equation}
\hat{\alpha} = n\left( \sum_{i=1}^n \log \frac{x_i}{x_{min}} \right),
\label{Hill}
\end{equation}

\noindent where $n$ is the number of data points $x_i$ used to estimate the power-law exponent. This estimator is known to be asymptotically normal and consistent in the limit of large number of data points $n$, and the standard error of the estimate $\hat{\alpha}$ can be calculated as: $\sigma = \hat{\alpha}/\sqrt{n}$. In this paper we analyze the power-law exponents in the tails of Bitcoin return distributions using both aforementioned methods, as a way to additionally authenticate our results. To estimate the cutoff point $x_{min}$ we employ a method proposed by Clauset et al. \cite{Clauset2009}, which relies on the Kolmogorov-Smirnov (KS) statistic as a measure of the distance between the empirical cdf ${Q}(x)$ and the fitted cdf $P(x)$:

\begin{equation}
D = \max_x \vert P(x) - Q(x) \vert,
\label{KS}
\end{equation}

\noindent which essentially equals the maximum distance between the two probability functions. The chosen cutoff point $x_{min}$ is then the one which minimizes the KS statistic for the region $x\geq x_{min}$. 

In practice it is easy to fit a power-law distribution and obtain an estimate for the exponent from data which do not necessarily conform to the power-law distribution. To test the power-law hypothesis in our data, we employ a resampling based method proposed by Preis et al. \cite{Preis2011}. First, the power-law exponent $\hat{\alpha}$ is estimated (using any chosen method) from the $n$ data points in the tails of the distribution, and the KS statistic $D^{emp.}$ for this fitted power-law distribution is calculated. Then, a large number $N$ of synthetic datasets is generated, all of which contain exactly $n$ points sampled from the power-law distribution with the exponent $\hat{\alpha}$. For each of these synthetic datasets the scaling exponent is estimated (using the same method) and the respective KS statistics $D_j$, $j = 1,...,N$ are calculated. If the deviation of the empirical data from the power-law fit $D^{emp.}$ is larger than a significant portion of the deviations of the synthetic datasets $D_j$, the power-law may not be a plausible explanation for the empirical distribution. Specifically, the fraction of the synthetic KS statistics $D_j$ larger than the empirical one $D^{emp.}$ is the p-value for the null hypothesis that the empirical data comes from a power-law distribution. Thus, if the p-value is small, the deviations of the empirical data cannot be attributed to stochastic effects alone and the power-law hypothesis is rejected. In our analyses, we use $N=1000$ generated synthetic datasets and a confidence level of $95\%$ to test the power-law hypothesis in the considered data.

\section{Results}

Since our data suggests that trading activity differs between exchanges and have evolved through time, we explore two issues: (i) whether there is an evolution of the scaling properties of Bitcoin returns through time, and (ii) do differences in market liquidity for considered exchanges impact these properties? To test these, we consider the $\Delta t = 5$ min returns and divide the data into $T = 15$ consecutive, non-overlapping 6-month periods, from August 2010 to February 2018. For each of these we estimate the scaling exponent from returns from all exchanges using the Hill estimator. We then fit a linear regression of the form: $\hat{\alpha}_t = \beta_0 + \beta_1 t$, where $t = 1,...,T$ denotes the 6-month period, and $\hat{\alpha}_t$ is the estimated exponent for that period. In Table \ref{time_reg_table} it is shown that although the coefficients $\beta_1$ associated with the temporal component are positive for both left (negative) and right (positive) tails, the corresponding t-tests cannot reject the null hypothesis $H_0: \beta_1 = 0$. Thus, even though a positive drift is observed, meaning that the estimated scaling exponents increase over time, the data does not provide sufficient evidence to support the hypothesis that time plays a significant role. However, it is possible that in future and over time, as the Bitcoin ecosystem evolves and as new data arrives, we witness a change in this behavior. Lastly, we report that the power-law hypothesis test based on generated synthetic datasets confirmed the null hypothesis (that the tail data follows a power-law) for 28 out of 30 tests (two tails $\times$ 15 periods), with a $95\%$ level of confidence, but when accounting for multiple tests by using the Bonferroni correction, none of the 30 null hypotheses were rejected.

\begin{table}[h]
\centering
\bgroup
\small
\setlength{\tabcolsep}{10pt}
\def\arraystretch{1.2}
\begin{tabular}{ r|c c|c c| }
\cline{2-5}
    &\multicolumn{2}{c|}{Positive tail}&\multicolumn{2}{c|}{Negative tail} \\ \cline{2-5}
	& $\beta_0$ &  $\beta_1$ &  $\beta_0$ &  $\beta_1$ \\ \hline
	Estimate & \textbf{2.442} & 0.034 & \textbf{2.471} & 0.031 \\
	p-value & $3.12 \cdot 10^{-7}$ & 0.215 & $1.65 \cdot 10^{-6}$ & 0.331 \\ \hline
\end{tabular}
\egroup
    \caption{Regression results for the estimated scaling exponents from $\Delta t = 5$ min data over time, including coefficient estimates and the p-values associated with the two-tailed t-test (coefficients significant at a 95\% level of confidence are shown in bold).}
    \label{time_reg_table}
\end{table}
  
In addition, we test whether market liquidity impacts the estimated scaling exponents. As before, we use the $T = 15$ non-overlapping consecutive half-year periods, but now we analyze each exchange separately, using $\Delta t = 5$ min data. For each combination of exchange and half-year period we estimate the scaling exponent using the Hill estimator and calculate the average traded volume in USD. Note that not all exchanges have data in all considered periods, thus we have a total of 45 data points for which we obtain the correlation between the estimated scaling exponents and trading volume: 0.22 for positive and 0.19 for negative tails. Although both are positive, implying that larger market liquidity is associated with larger power-law exponents (i.e. thinner tails), the respective p-values based on the t-test are 0.13 and 0.19, indicating that the evidence from data is not strong enough to reject the null hypothesis. Therefore, even if the effects exists, it seems not to be critical or decisive in the considered exchanges, and can be left to further analyses as new data arrives. Based on these results we choose not to pursue further modelling of these rather intricate patterns and, following the principle of Occam's razor, in the following we consider data from all exchanges and the entire 2010-2018 time period.

\begin{figure}[h!]
\centering
\begin{subfigure}{.5\textwidth}
  \centering
  \includegraphics[width=0.9\linewidth]{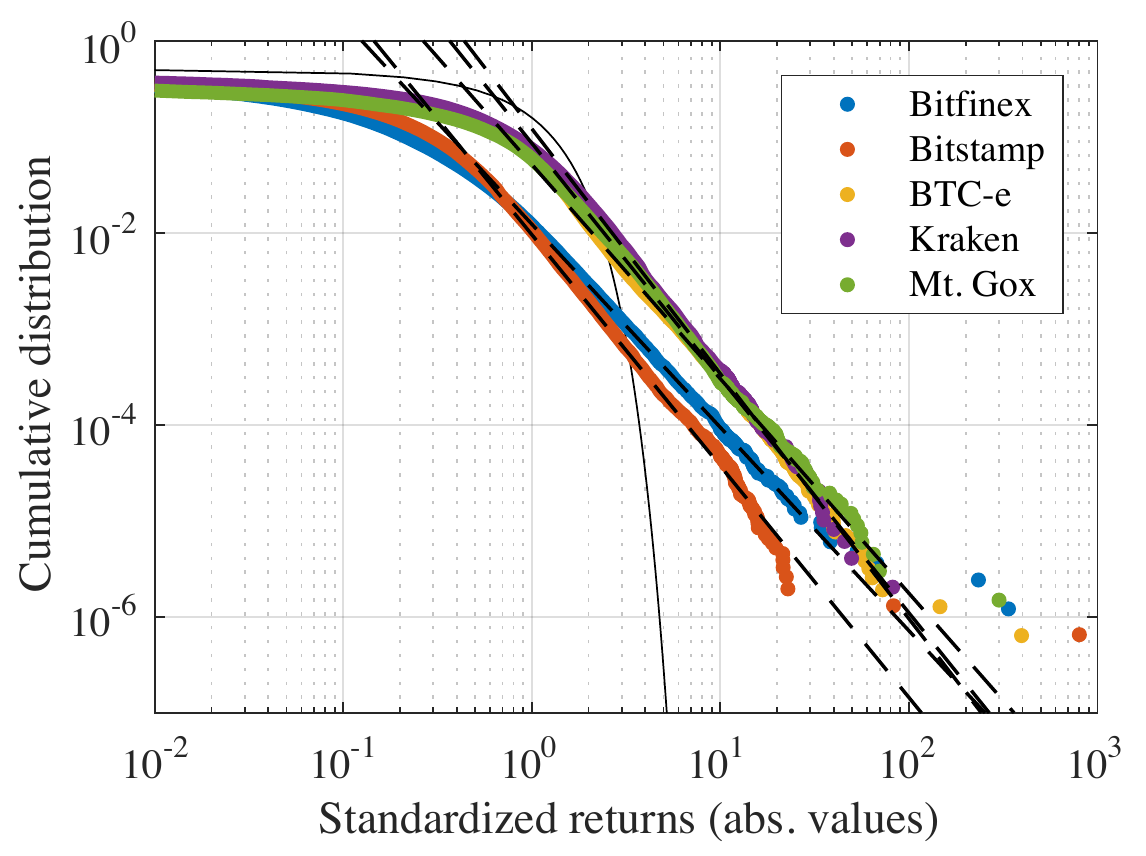}
  \label{}
\end{subfigure}%
\begin{subfigure}{.5\textwidth}
  \centering
  \includegraphics[width=0.9\linewidth]{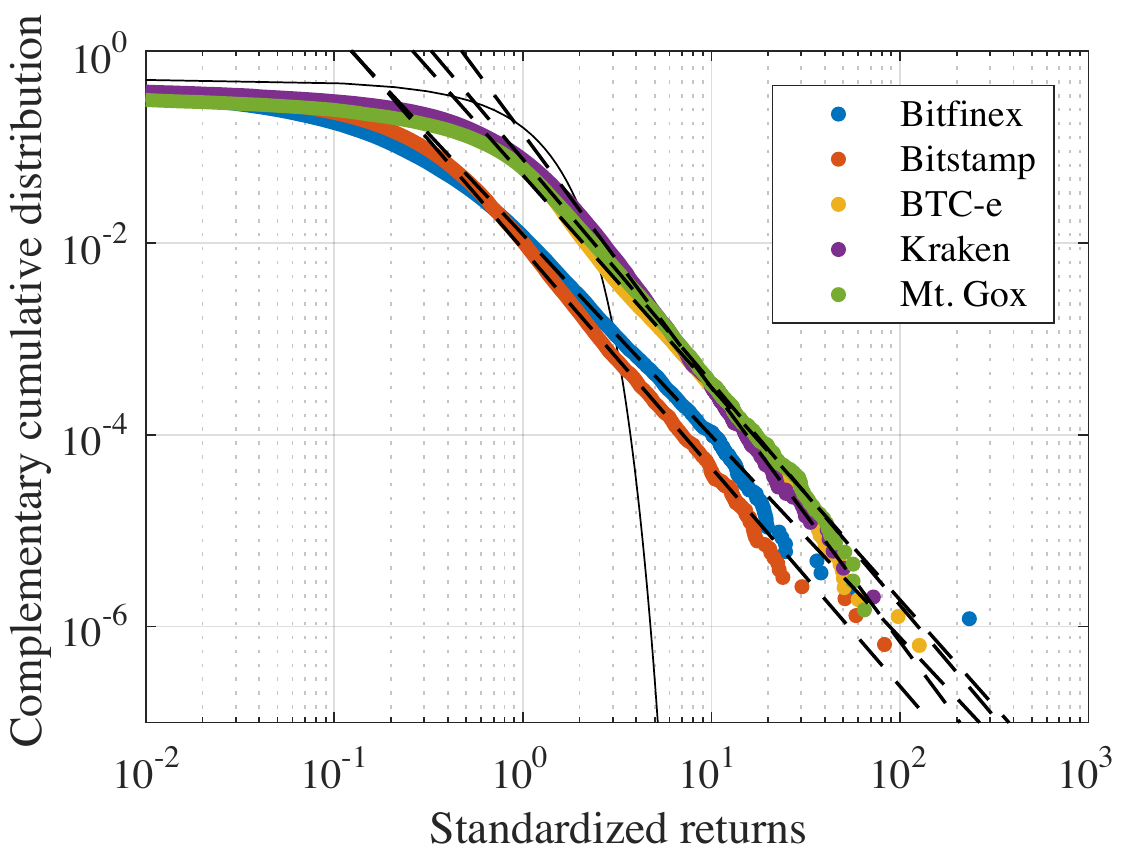}
  \label{}
\end{subfigure}
\caption{Negative (left) and Positive (right) tails of the cumulative distribution for the Bitcoin returns on the considered exchanges and the time scale of $\Delta t = 1$ min. The black dashed lines correspond to the fitted power-law distributions for the return tails of the considered exchanges and the black full line is the cumulative distribution function of the standard normal $\mathcal{N}(0,1)$.}
\label{tails_all}
\end{figure}

First, we analyze the empirical distributions of (normalized) Bitcoin returns observed from different exchanges at short time scales: $\Delta t = 1$ min, $5$ min, and $10$ min. For these very short time intervals we can obtain the largest amount of data and thus analyze data from each exchange separately. Figure \ref{tails_all} displays the complementary cumulative distributions for positive tails and the cumulative distribution for negative tails of standardized returns for the considered exchanges. 

\noindent The distributions shown in Figure \ref{tails_all} vary slightly between different exchanges, but their tails seem to decay in a similar fashion, and evidently much slower than the tails of the Gaussian distribution. This suggests that although differences in the dynamics of prices in these exchanges exist (which is expected, since data from different exchanges corresponds to different periods in the past), the tail behavior seems to be a common generality. 

We apply both the log-log regression fit and the Hill estimator, with the cutoff point $x_{min}$ estimated as the one which minimizes the KS statistic, to the Bitcoin returns over high-frequency time intervals of $\Delta t = 1$min, 5min, and 10min on each considered exchange separately. The results are given in Table \ref{high_freq_exp_pos} for the positive and Table \ref{high_freq_exp_neg} for the negative tails of the empirical distributions. We also report the standard errors obtained for the Hill estimator and the p-values associated with the performed hypothesis tests for the power-law shape of the distribution tails in brackets below each estimate. 

\begin{table}[h!]
\centering
\small
\bgroup
\def\arraystretch{1.8}
\begin{adjustbox}{center}
\begin{tabular}{ |c|c|ccccc| }
\hline
     \multirow{2}{*}{$\Delta t$}&\multirow{2}{*}{Est.}&\multicolumn{5}{c}{Exchange} \vline \\ 
     &&{Bitfinex} & {Bitstamp} & {BTC-e} & {Kraken} & {Mt. Gox} \\ \hline

\multirow{2}{*}{\rotatebox[origin=c]{90}{1 min}}&\footnotesize{reg.} & \stackanchor{2.10}{\scriptsize{(0.545)}} & \stackanchor{2.28}{\scriptsize{(0.380)}} & \stackanchor{2.21}{\scriptsize{(0.165)}} & \stackanchor{2.65}{\scriptsize{(0.941)}} & \stackanchor{2.32}{\scriptsize{(0.158)}}\\
&\footnotesize{Hill} & \stackanchor{$2.15\pm0.11$}{\scriptsize{(0.931)}} & \stackanchor{$2.25\pm0.10$}{\scriptsize{(0.723)}} & \stackanchor{$2.14\pm0.03$}{\scriptsize{(0.114)}} & \stackanchor{$2.64\pm0.06$}{\scriptsize{(0.844)}} & \stackanchor{$2.41\pm0.06$}{\scriptsize{(0.840)}}\\[1ex]\hline

\multirow{2}{*}{\rotatebox[origin=c]{90}{5 min}}&\footnotesize{reg.} & \stackanchor{2.11}{\scriptsize{(0.775)}} & \stackanchor{2.23}{\scriptsize{(0.575)}} & \stackanchor{2.07}{\scriptsize{(0.000)}} & \stackanchor{2.49}{\scriptsize{(0.766)}} & \stackanchor{2.32}{\scriptsize{(0.256)}}\\
&\footnotesize{Hill} & \stackanchor{$2.11\pm0.09$}{\scriptsize{(0.535)}} & \stackanchor{$2.14\pm0.07$}{\scriptsize{(0.855)}} & \stackanchor{$2.08\pm0.05$}{\scriptsize{(0.008)}} & \stackanchor{$2.42\pm0.09$}{\scriptsize{(0.434)}} & \stackanchor{$2.19\pm0.07$}{\scriptsize{(0.059)}}\\[1ex]\hline

\multirow{2}{*}{\rotatebox[origin=c]{90}{10 min}}&\footnotesize{reg.} & \stackanchor{2.24}{\scriptsize{(0.287)}} & \stackanchor{2.26}{\scriptsize{(0.791)}} & \stackanchor{2.37}{\scriptsize{(0.187)}} & \stackanchor{2.57}{\scriptsize{(0.907)}} & \stackanchor{2.33}{\scriptsize{(0.074)}}\\
&\footnotesize{Hill} & \stackanchor{$2.22\pm0.10$}{\scriptsize{(0.805)}} & \stackanchor{$2.24\pm0.07$}{\scriptsize{(0.311)}} & \stackanchor{$2.27\pm0.08$}{\scriptsize{(0.813)}} & \stackanchor{$2.49\pm0.12$}{\scriptsize{(0.880)}} & \stackanchor{$2.28\pm0.12$}{\scriptsize{(0.929)}}\\[1ex]\hline

\end{tabular}
\end{adjustbox}
\egroup

    \caption{Estimated power-law exponents $\alpha$ for the positive tail of the empirical distributions of Bitcoin returns over high frequency time intervals on each considered exchange. The estimated scaling exponenst are just beyond the Levy alpha-stable regime characterized by $\alpha$ between 0 and 2.}
    \label{high_freq_exp_pos}
    
\end{table}

\begin{table}[h!]
\centering
\small
\bgroup
\def\arraystretch{1.8}
\begin{adjustbox}{center}
\begin{tabular}{ |c|c|ccccc| }
\hline
     \multirow{2}{*}{$\Delta t$}&\multirow{2}{*}{Est.}&\multicolumn{5}{c}{Exchange} \vline \\ 
     &&{Bitfinex} & {Bitstamp} & {BTC-e} & {Kraken} & {Mt. Gox} \\ \hline

\multirow{2}{*}{\rotatebox[origin=c]{90}{1 min}}&\footnotesize{reg.} & \stackanchor{2.12}{\scriptsize{(0.842)}} & \stackanchor{2.41}{\scriptsize{(0.750)}} & \stackanchor{2.23}{\scriptsize{(0.136)}} & \stackanchor{2.55}{\scriptsize{(0.251)}} & \stackanchor{2.45}{\scriptsize{(0.542)}}\\
&\footnotesize{Hill} & \stackanchor{$2.08\pm0.11$}{\scriptsize{(0.963)}} & \stackanchor{$2.30\pm0.09$}{\scriptsize{(0.331)}} & \stackanchor{$2.17\pm0.03$}{\scriptsize{(0.057)}} & \stackanchor{$2.44\pm0.06$}{\scriptsize{(0.708)}} & \stackanchor{$2.48\pm0.05$}{\scriptsize{(0.850)}}\\[1ex]\hline

\multirow{2}{*}{\rotatebox[origin=c]{90}{5 min}}&\footnotesize{reg.} & \stackanchor{2.23}{\scriptsize{(0.847)}} & \stackanchor{2.34}{\scriptsize{(0.648)}} & \stackanchor{2.09}{\scriptsize{(0.001)}} & \stackanchor{2.84}{\scriptsize{(0.895)}} & \stackanchor{2.42}{\scriptsize{(0.976)}}\\
&\footnotesize{Hill} & \stackanchor{$2.18\pm0.09$}{\scriptsize{(0.891)}} & \stackanchor{$2.31\pm0.05$}{\scriptsize{(0.926)}} & \stackanchor{$2.03\pm0.05$}{\scriptsize{(0.004)}} & \stackanchor{$2.83\pm0.12$}{\scriptsize{(0.820)}} & \stackanchor{$2.41\pm0.08$}{\scriptsize{(0.970)}}\\[1ex]\hline

\multirow{2}{*}{\rotatebox[origin=c]{90}{10 min}}&\footnotesize{reg.} & \stackanchor{2.05}{\scriptsize{(0.019)}} & \stackanchor{2.27}{\scriptsize{(0.722)}} & \stackanchor{2.07}{\scriptsize{(0.000)}} & \stackanchor{2.68}{\scriptsize{(0.860)}} & \stackanchor{2.32}{\scriptsize{(0.428)}}\\
&\footnotesize{Hill} & \stackanchor{$2.00\pm0.07$}{\scriptsize{(0.064)}} & \stackanchor{$2.25\pm0.06$}{\scriptsize{(0.725)}} & \stackanchor{$2.23\pm0.07$}{\scriptsize{(0.021)}} & \stackanchor{$2.72\pm0.15$}{\scriptsize{(0.737)}} & \stackanchor{$2.31\pm0.11$}{\scriptsize{(0.955)}}\\[1ex]\hline

\end{tabular}
\end{adjustbox}
\egroup

    \caption{Estimated power-law exponents $\alpha$ for the positive tail of the empirical distributions of Bitcoin returns over high frequency time intervals on each considered exchange. The estimated scaling exponenst are just beyond the Levy alpha-stable regime characterized by $\alpha$ between 0 and 2.}
    \label{high_freq_exp_neg}
    
\end{table}

It is evident that, although some variations exist between different exchanges, there is a general concordance between the estimates, suggesting that the power-law exponent for intraday Bitcoin returns is within the range of $2 <\alpha< 2.5$. In addition, it is important to note that the coefficients $\beta_0$ from Table \ref{time_reg_table}, estimated for 5 min returns, also suggest that the scaling exponent might be around 2.5, confirming these results. Another interesting result is the fact that no obvious asymmetries between positive and negative tails of the considered distributions are found. Furthermore, a vast majority of the performed statistical tests could not reject the null hypothesis that the tails follow a power-law at a 95\% level of confidence, with the exception of a few estimates for the BTC-e exchange (although we did not adjust for multiple tests, in which case some of these would be corrected). Most notably, the estimated power-law exponents are outside the Levy-stable region $0<\alpha<2$ implies the existence of a finite second moment for Bitcoin return distributions, which is important for risk modelling scenarios.

\begin{figure}[h!]
\centering
\includegraphics[width=0.65\linewidth]{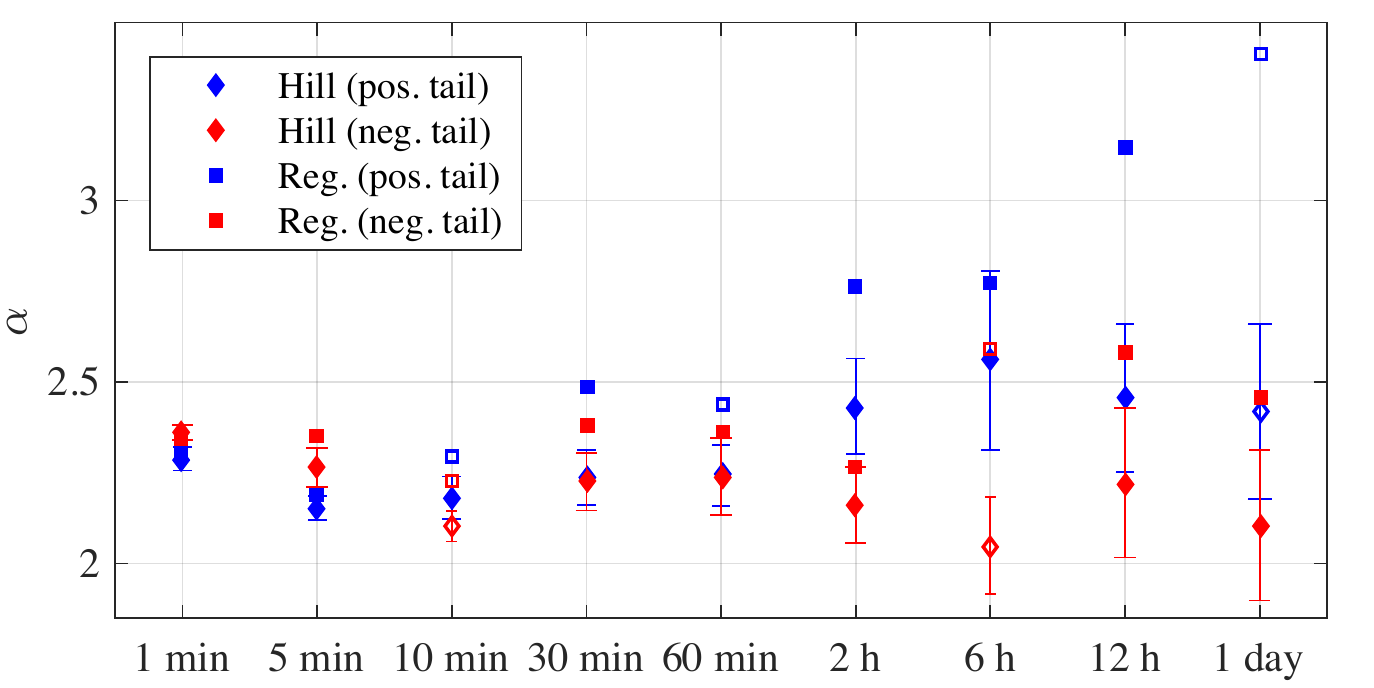}
\caption{Estimates of the power-law exponents for the positive and negative tails of Bitcoin returns over multiple time intervals. Fully colored points in the graph represent estimates for which the power-law hypothesis could not be rejected, whereas the empty points correspond to those for which the null hypothesis was rejected at a 95\% level of confidence (without correcting for multiple tests).}
\label{all_exp}
\end{figure}

Next, we analyze the tails of empirical distributions of returns over multiple larger time intervals. Motivated by the results in the previous section, where similar behavior was found for Bitcoin returns across all the considered exchanges, we aggregate the returns for all the exchanges and consider one aggregate distribution. By doing so, we are able to acquire much more data and thus consider tails of empirical distributions for larger time scales, whereas data from single exchanges would be insufficient for tail estimation. We report the tail exponent estimates for the empirical distributions of returns on various time intervals in Figure \ref{all_exp} for both positive negative tails. Most of the estimated power-law exponents can be seen to fit into the $2 <\alpha< 2.5$ range. These results generally confirm the implications suggested by the estimated power-law exponents for high frequency return distributions, but seem to fluctuate strongly towards longer time intervals between different estimators and considered tails, most likely due to scarcity of data. Thus, an investigation of these scaling properties of larger time interval returns, such as those performed on decade-long stock price datasets, is not yet possible for this developing asset class. However, our results present compelling evidence about the general nature of the scaling properties of extreme price fluctuations from multiple Bitcoin exchanges and the data compiled so far. A possible explanation for the heavier tails observed in cryptocurrency markets, as opposed to those in stock markets, might be a relatively different structure of cryptocurrency investors than that of traditional investors, as suggested by previous studies \cite{Kristoufek2013,Garcia2014}. The observed bubbles testify to the level of feedback and herding effects in these markets. In addition, if we think of the power-law behavior as a consequence of investors’ poor estimation of the true asset value \cite{Kostanjcar2013} - it is only logical that the observed tails are heavier, since the intrinsic value for cryptocurrencies is especially hard to determine. However, a more detailed study of the behavioral issues in these markets is necessary to reach definite conclusions.

\section{Conclusion}
We have analyzed Bitcoin trade data from multiple digital exchanges (Mt. Gox, BTC-e, Bitstamp, Bitfinex, and Kraken) on a time period from 2010 to 2017 comprising a total of over 80 million trades, focusing on the distribution tails of returns over different time intervals. We find evidence of a power-law decay in the tails of empirical cumulative distributions of price fluctuations over multiple time intervals, ranging from $\Delta t = 1$ min to $\Delta t = 1$ day. We employ two methods to estimate the scaling exponents and report that Bitcoin returns exhibit heavy tails with $2< \alpha < 2.5$, as opposed to the inverse cubic law found for stock returns. Moreover, we find that this behavior holds across multiple exchanges and for multiple tine intervals, suggesting that this is a universal phenomenon, similar to that found in other financial markets. Our results suggest that in addition to a much higher variance, Bitcoin returns also exhibit heavier tails than stock returns, meaning that extreme price fluctuations are considerably more frequent. In addition, our results also imply the existence of a finite second moment for Bitcoin price fluctuations implying the central limit theorem and converging to a Gaussian limit. These findings provide a fundamental basis for the usage of standard financial theories and covariance-based techniques in risk management and portfolio optimization scenarios.

\section{Acknowledgment}
This work has been supported in part by the Croatian Science Foundation under the project 5349. The authors would like to thank the reviewers for their comments which helped improve the quality of the paper.

\section{Bibliography}
\bibliographystyle{ieeetr}
\bibliography{btc_scaling}

\end{document}